\documentstyle[aps,prl,twocolumn,psfig]{revtex}
\begin{document}

\twocolumn[\hsize\textwidth\columnwidth\hsize\csname@twocolumnfalse\endcsname
\title{\bf Mesoscopic proximity effect in double barrier Superconductor/Normal Metal junctions}
\author{D. Quirion, C. Hoffmann, F. Lefloch and M. Sanquer}
\address{D\'epartement de Recherche Fondamentale sur la Mati\`ere Condens\'ee/SPSMS/LCP \\ CEA Grenoble \\ 17, rue des Martyrs \\ 38054 Grenoble Cedex, FRANCE.}

\maketitle

\begin{abstract}

We report transport measurements down to $T=60mK$ of SININ and SNIN structures in the diffusive limit. We fabricated $Al-AlOx/Cu/AlOx/Cu$ (SININ) and $Al/Cu/AlOx/Cu$ (SNIN) vertical junctions. For the first time, a zero bias anomaly was observed in a metallic SININ structure. We attribute this peak of conductance to coherent multi-reflections of electrons between the two tunnel barriers. This conductance maximum is quantitatively fitted by the relevant theory of mesoscopic SININ structures. When the barrier at the SN interface is removed (SNIN structure), we observe a peak of conductance at finite voltage accompagnied by an excess of sub-gap conductance.
\end{abstract}]

\par Transport through SIN junction is a very long known problem for strong barrier which leads to the spectroscopy of the superconductor \cite{Giaever60}. However, when the resistance of the normal metal is comparable to the resistance of the barrier, a novel quantum phenomena, called reflectionless tunneling, takes place. Experimentally, it is measured as a peak at zero voltage in the differential conductance characteristics of the SIN junction. This zero voltage maximum conductance comes from coherent multiple Andreev reflections produced by the backscattering of electrons towards the SIN interface induced by disorder in the normal electrode \cite{vanWees92,Yip95,Volkov94,Hekking94}. This effect was first observed by Kastalsky et al. \cite{Kastalsky91} in Nb/InGaAs contact. Since then, reflectionless tunneling was mainly measured in superconductor/semiconductor junctions with a moderate transparent interface due to the annealed Schottky barrier \cite{Bakker94,Magnee94,Quirion00,Giazotto01}. Very few results were obtained in metallic SIN junction because a normal metal is not disordered enough to retroreflect electron towards the SIN interface and the oxide barrier is too opaque. 

\par It was proposed to compensate the lack of disorder in the normal metal by a second tunnel barrier at some distance $d$ from the SIN interface \cite{Melsen94,Volkov93,Lesovik97}. The resulting SININ structure would be the electronic analog of the Fabry-Perot effect in optical cavities. Coherent multiple reflection between the two tunnel barrier enhance the conductance at low energy and lead to a peak of conductance, equivalent to the reflectionless tunneling anomaly in semiconductor/superconductor contacts. The peak of conductance is not sensitive to elastic diffusion inside the device, because of the electron-hole conjugaison induced by the Andreev mirror \cite{Beenakker97}. However, if the electron and the Andreev-reflected hole have not the same energy (because of finite temperature or voltage), or in presence of a magnetic flux, a phase accumulates between the two Andreev reflections, breaking the coherent addition of electronic amplitudes: the effect is then destroyed. Theories also predict a crossover from this zero bias anomaly to a finite voltage maximum conductance when the transparency of the barrier at the SIN interface is higher than the transparency of the second barrier. This finite bias anomaly is due to the opening of a gap in normal layer between the two tunnel barrier. Such a gap was observed in N/S bilayers: Vinet et al. \cite{Vinet01} and Moussy et al. \cite{Moussy01} studied diffusive $Nb/Au$ bilayers by STM spectroscopy and measured an induced (mini)gap in the normal metal due to the proximity effect. Such a behaviour was also observed by Gu\'eron et al. \cite{Gueron96} with tunnel junctions deposited on a copper wire in good contact with a superconductor (aluminum).

\par In this work we report the first observation of the zero voltage maximum conductance in a metallic SININ junction, as well as the observation of a finite voltage maximum conductance in its SNIN counterpart.

\par
\begin{figure}
	\begin{center}
	\psfig{figure=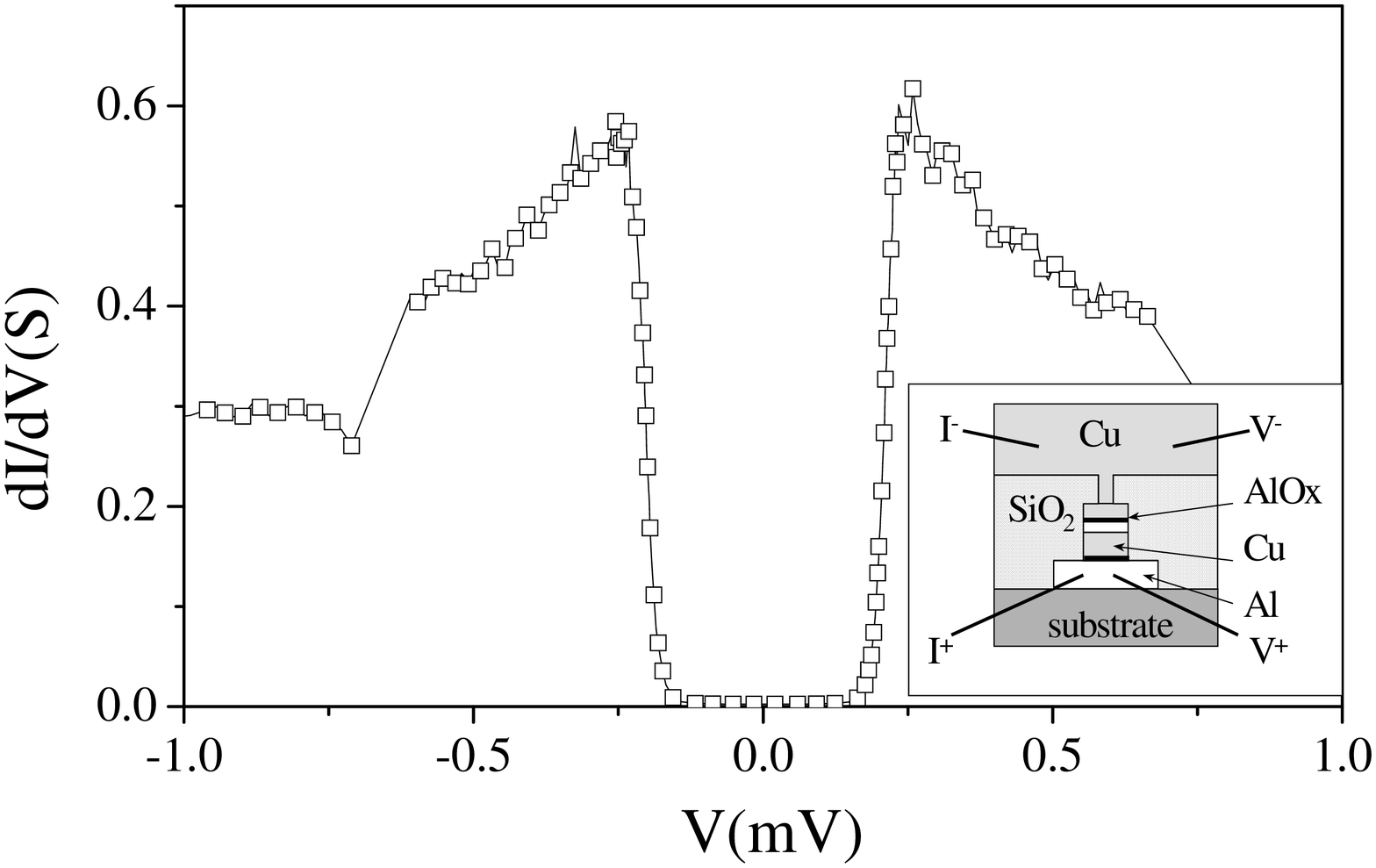,width=87mm}
	\caption{Differential conductance versus voltage of a $10\times 10\mu m^{2}$ SIN junction at T=60mK. The sudden drop at $V \simeq 0.7 V$ is due to a critical current transition effect in the aluminum bank. No maximum of conductance is seen at low voltage. Inset: Schematic cross section the vertical SININ structure.}
	\label{fig:g-vsin}
	\end{center}
\end{figure}

 The SININ samples were fabricated by depositing in situ an $Al/Cu/Al-AlOx/Cu$ multilayer in a DC magnetron sputtering machine with a base pressure of $10^{-7}mbar$. First, a sequence $Al(150nm)-AlOx/Cu(d)/Al(10nm)$ was sputtered on a $Si/SiO_{2}$ 3' wafer. The base electrode of aluminum was oxidized with an exposure of 2mTorr during one minute. To form the NIN (Cu-AlOx-Cu) barrier, the thin $Al(10nm)$ layer was oxidized  without breaking the vacuum under a controlled atmosphere of oxygen ($p_{ox}=1mbar$) during one hour. This aluminum layer is not fully oxidized but the remaining bilayer (Cu-Al) is not superconducting \cite{deGennes64}. We want to stress the great asymmetry in oxidation conditions between the two barriers. They are of very different nature (thick aluminum electrode oxidized vs thin oxidized aluminum layer deposited on top of a copper layer). Consequently, very different oxidation conditions are needed to obtain comparable transparencies (see below). The multilayer deposition was completed by a 50nm copper protection layer. Then, the base electrode and the junctions were defined in a two steps optical lithography and dry etching procedure. The sides of the junctions were insulated by deposition and lift-off of silicon dioxide ($300nm$). Finally, a copper counter-electrode ($500nm$ thick) was deposited (see insert of figure \ref{fig:g-vsin}). The junction areas $S$ range from $2\times2$ to $30\times30 \mu m^{2}$. We estimated the diffusion coefficient to $D=64cm.s^{-1}$ and the mean free path to $\ell=12nm$ in copper, so that the copper layer is  in the diffusive limit. The SNIN samples were fabricated using the same procedure, except the base electrode of aluminum was not oxidized.

\par Before considering   SININ or SNIN junctions, we  measured  a simple $10\times10 \mu m^{2}$ $Al-AlOx/Cu$ (SIN) junction fabricated with an oxidation of 1mbar during one hour (see figure \ref{fig:g-vsin}). From the value of the normal conductance $G_{NN}=0.36S$, we estimate its transmission coefficient $\Gamma_{SIN}=G_{NN} (\lambda_{F}/2)^{2}/G_{Q}S\simeq 2\, 10^{-6}$, with $G_{Q}=(12885\Omega)^{-1}$ the quantum conductance and $\lambda_{F}=0.45nm$ the Fermi wave-length in copper. The normal conductance is in good agreement with values measured by Kleinsasser et al.\cite{Kleinsasser95}. They estimate the barrier resistance $R_{b}$ as a function of exposure $E$ (oxidation duration times oxygen pressure): $R_{b}(\Omega.\mu m^{2}) \simeq 2[E(Pa.s)]^{0.4} \simeq 330\Omega.\mu m^{2}$ in our exposure conditions, in good agreement with the barrier resistance $100/0.36=280\Omega.\mu m^{2}$. Note that, due to the large area of the junction, $G_{NN}$ is much larger than in others measurements \cite{Vinet01,Moussy01,Gueron96}, that limits possible Coulomb blockade effects. 

\par As (uncoherent) Andreev reflection is the only transport process available below the superconducting gap, one expects the ratio subgap conductance to normal conductance to be here $G_{subgap}/G_{NN}=2\Gamma_{SIN}=4\, 10^{-6}$. However, we measure $G_{subgap}/G_{NN}=7\, 10^{-3}$. This can not be explained by thermal excitations, as they are exponentially small. Such a ratio ($\sim 10^{-3}$) is consistent with other experimental studies \cite{Kleinsasser95,Pothier94} and may be due to inhomogeneities in the barrier. Another way to reconcile the observed relatively small $G_{subgap}/G_{NN}$ ratio with theory is to introduce a small finite lifetime for quasiparticles in the superconductor $\Gamma_{s} \simeq 0.0067 \Delta$ \cite{Dynes78}. The conductance-voltage characteristics of the SIN junction follows the usual Hamiltonian tunnel behaviour and reproduces reasonably the BCS density of states of aluminum ($\Delta=205\mu eV$). The critical temperature of the aluminum banks is $T=1.5K$.

\begin{figure}
	\begin{center}
	\psfig{figure=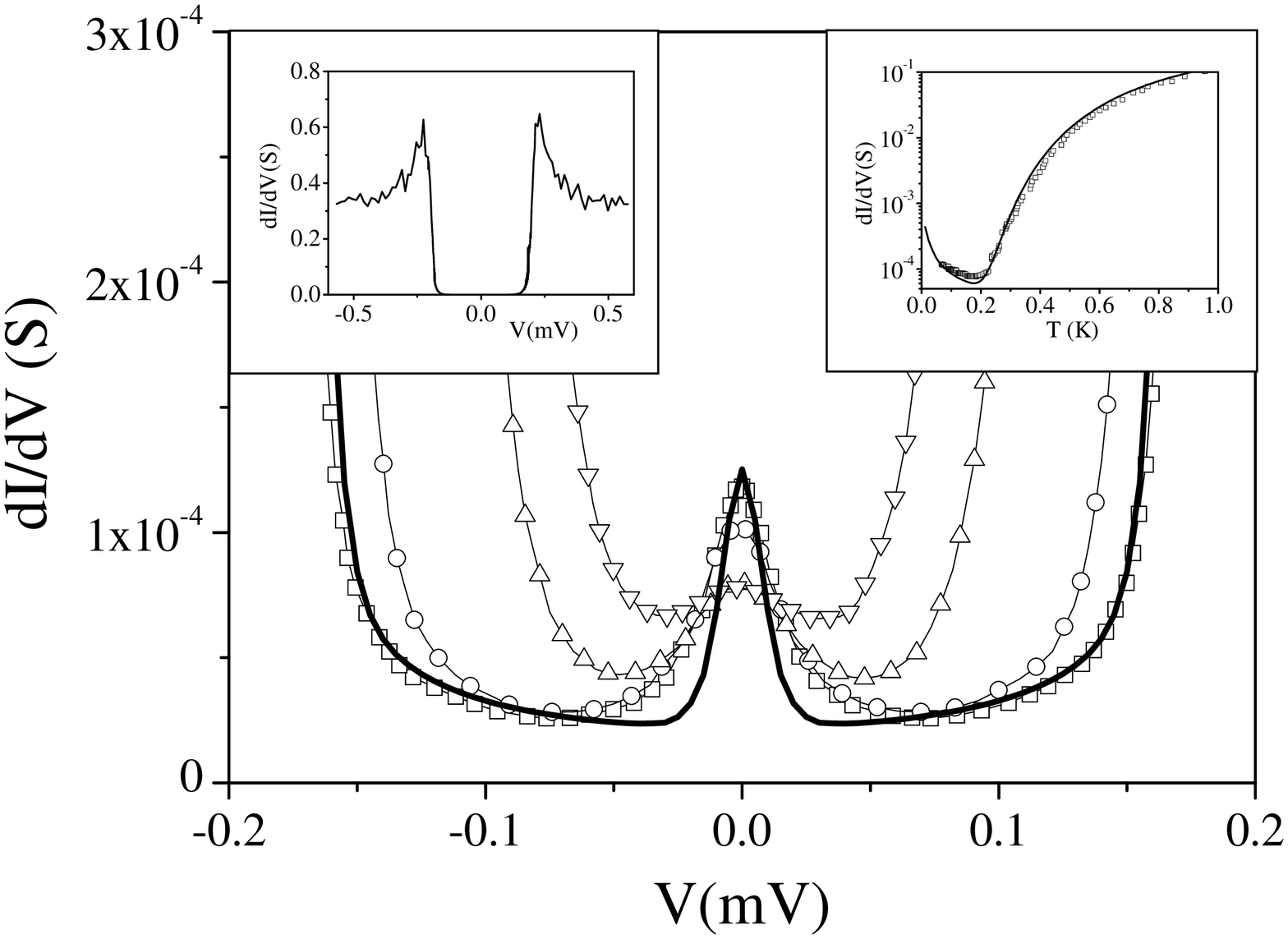,width=87mm}
	\caption{Conductance versus voltage characteristics of a $20\times 20\mu m^{2}$ SININ sample at various temperature $T=60, 90, 150, 180$ and $225mK$ from bottom to top. The conductance steadily increases at low voltage ($V<50\mu eV$) below $T \simeq 200mK$. The solid line represents the theoretical fit using the following parameters: $T=60mK$, $\Delta=210\mu eV$, $\Gamma_{SIN}=5.10^{-7}$, $\Gamma_{NIN}=2.10^{-6}$, $\Gamma_{s}=4.10^{-5}\Delta$ and $G_{NN}=0.23S$. Inset: Temperature dependence of the zero voltage conductance of the same sample. The solid line is the theoretical fit with the parameters noted above.}
	\label{fig:g-tv0}
	\end{center}
\end{figure}

\par We then measured a $20\times 20\mu m^{2}$ SININ structure. At large voltage, its differential conductance-voltage characteristics is close to the expected SIN-like characteristics (see figure \ref{fig:g-vsin}). Figure \ref{fig:g-tv0} shows the differential conductance-voltage characteristics of this SININ structure at various temperature at low voltage and the temperature dependence of the zero voltage conductance. We observe a peak of conductance around zero voltage we attributed to coherent multiple reflections between the two alumina barriers. This zero voltage maximum conductance is destroyed at temperature above $T \simeq 200mK$, $V \simeq 50\mu V$ and $H \simeq 10^{-2}T$ (data not shown). We use the theory of Volkov et al. \cite{Volkov93} for the diffusive SININ structure to fit our data. We obtain an excellent agreement using the following parameters: $T=60mK$, $\Delta=210\mu eV$, $\Gamma_{SIN}=5.10^{-7}$, $\Gamma_{NIN}=2.10^{-6}$, $\Gamma_{s}=4.10^{-5}\Delta$ and $G_{NN}=0.23S$. This normal conductance leads to a effective transparency of the structure of $\Gamma_{SININ}=G_{NN} (\lambda_{F}/2)^{2}/G_{Q}S\simeq 4\, 10^{-7}$, in very good agreement with the obtained transparencies of the barriers: $(\Gamma_{SIN}^{-1}+\Gamma_{NIN}^{-1})^{-1} \simeq 4\, 10^{-7}$. We also note that the transparencies of the barriers are similar  although the oxidation conditions are very different. This points out the very different nature of the two alumina barriers.

\begin{figure}
	\begin{center}
	\psfig{figure=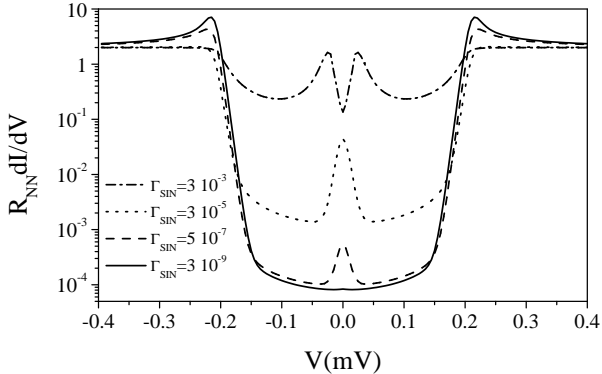,width=87mm}
	\caption{Three types of behaviors for the differential conductance versus voltage, as predicted by the theory \protect\cite{Volkov93} for an SININ structure when the  transparency $\Gamma_{NIN}$ of the NIN barrier is varied. Other parameters are kept constant: $T=60mK$, $d=30nm$, $D=60cm^{2}s^{-1}$ and $\Gamma_{SIN}=5.10^{-7}$. For a small NIN barrier (large transparency $\Gamma_{NIN}$), the conductance is flat in voltage (at low voltage) as in figure \ref{fig:g-vsin} for an SIN junction. For comparable barriers $\Gamma_{NIN} \simeq \Gamma_{SIN}$, a zero voltage maximum conductance is obtained as in figure \ref{fig:g-tv0} (the fit is the same as presented on figure \ref{fig:g-tv0}). Finally, for a small interface barrier SIN  ($\Gamma_{NIN} \ll \Gamma_{SIN}$) a finite voltage maximum conductance is obtained as in figure  \ref{fig:g-v2} for the SNIN case. In this last case nevertheless, the value of the subgap conductance (in particular the excess of subgap conductance) cannot be reconciled with the theory.}
	\label{fig:theory}
	\end{center}
\end{figure}

\par Figure \ref{fig:theory} shows the results of the Volkov's theory for an SININ diffusive junction when the  transparency $\Gamma_{NIN}$ of the NIN barrier is varied: if $\Gamma_{NIN}$ is too large as compared to $\Gamma_{SIN}$, the zero voltage conductance maximum is not observable. In this limit, the backscattering towards the SIN interface is not large enough to produce the effect in our experimental conditions. This is what we observe in the SIN case (see figure \ref{fig:g-vsin}). On the contrary if $\Gamma_{NIN}$ is small as compared to $\Gamma_{SIN}$, the zero voltage maximum evolves into a finite voltage maximum conductance. In order to observe this crossover between a zero and a finite voltage maximum, we considered a good SN interface case (large $\Gamma_{SIN}$), i.e. the generic SNIN case.

\par
\begin{figure}
	\begin{center}
	\psfig{figure=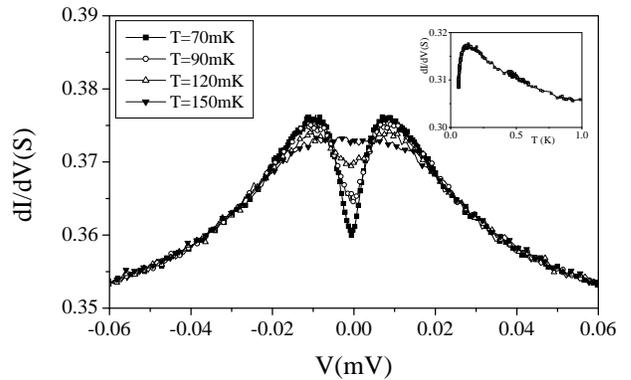,width=87mm}
	\caption{Differential conductance versus voltage  at various temperatures. A finite bias anomaly is measured at $11\mu V$. Note the excess of subgap conductance as in mesoscopic SN junctions, contrarily to the predictions of the tunnel theory for SNIN junctions. Inset: temperature dependence of the zero voltage conductance in the SNIN sample. There is a maximum of conductance at T=140 mK, and a decrease of the conductance at lower temperature. We note the good correspondance between voltage and temperature scales.}
	\label{fig:g-v2}
	\end{center}
\end{figure}

  Figure \ref{fig:g-v2} shows the differential conductance versus voltage at various temperatures for the SNIN junction and the temperature dependence of the zero voltage conductance: one observes indeed a finite voltage maximum conductance as expected. The differential conductance shows a dip below $25\mu V$ at low temperature. The low voltage feature decreases with increasing temperature and vanishes around $T\simeq 140mK$. We also observe that the maximum conductance is destroyed by a magnetic field of $10^{-2}T$ (data not shown). The conductance peak is around $11\mu V$ and does not move significantly with temperature. We note a good agreement between voltage ($V_{c}=11 \mu V$) and temperature scales ($T_{c}=140mK\simeq 12\mu eV/k_{B}$). This accordance has already been noticed in the reentrance context \cite{Courtois99,Petrashov98} or in other mesoscopic devices exhibiting coherent phenomena \cite{Quirion00}. It was explained as heating effect in the normal reservoir. Theories do not predict such an accordance if an equilibrium distribution function is supposed in the normal reservoir, mainly because of the asymmetric role of voltage and temperature in the Fermi function. Usually, larger voltage scales are expected but not observed. 

\par We cannot obtain a quantitative accordance with the theory, in particular because we observe an excess of subgap-conductance, i.e. the integrated differential conductance is larger than $G_{NN} \times \Delta$, where $G_{N}$ is the normal conductance above the gap $\Delta$. This is in contradiction with a simple tunnel spectroscopy, the dip at low voltage being the minigap in the N/S bilayer. Such a measurement would be the product of the densities of states on each sides of the alumina barrier and would obey an area conservation law, corresponding to the conservation of the number of electrons. On the other hand,  this excess is very similar to the reentrance observed by Charlat et al. \cite{Charlat96} in mesoscopic $Al/Cu$ contact, although the system is completely different.

\par A possible explanation to reconcile our observation with the spectroscopy of the minigap could be the presence of heating at finite voltage. Indeed, there is no more conservation argument for the sub-gap conductance if the effective temperature for Fermi distribution of electrons varies with the voltage. As noted before the observed coincidence between the temperature and the voltage at which the conductance is maximum is an indication that heating effects are probably important. We do not know of any model to take into account this effect in the context of an SNIN system.

\par Qualitatively, shifting the tunnel barrier from the  superconducting interface (SNIN geometry) leads to a very large increase of the sub-gap conductance as compared to SIN. Our explanation is based on the existence of Andreev states in the SN bilayer which carry a large sub-gap current. This current can even exceed the normal current value at resonances, as was noted in the ballistic case \cite{Rowell66,Hahn85,Fauchere98}.

\par Even if we cannot compare quantitatively the absolute value of the subgap conductance with the theory, we note that within the formalism of Volkov et al. \cite{Volkov93}, the differential conductance is maximum at a voltage given by $eV \simeq {hD\over L^{2}} \times {\Gamma_{SIN} d \over \ell}$ (see figure \ref{fig:theory}). If we suppose no heating effect to shift the maximum of conductance, parameters of our samples ($V= 11 \mu V$, $\ell =12nm$, $d=30nm$ and $D=60cm^{2}.s^{-1}$) lead to an estimation of $\Gamma_{SIN} \simeq 2.10^{-3}$, an intermediate value between a clean interface and a tunnel barrier, suggesting that the interface is disordered. Unfortunately, a direct measurement of this factor is very difficult.

\par We have fabricated $Al-AlOx/Cu/AlOx/Cu$ (SININ) and $Al/Cu/AlOx/Cu$ (SNIN) structures in the diffusive regime. For the first time we observe the zero voltage maximum conductance in a double barrier SININ system, which is the analog of the Fabry-Perot experiment with a superconducting mirror. Furthermore, by increasing the SN interface transparency, we observe the crossover between the zero voltage and the finite voltage maximum conductance. In SNIN, we measured a sub-gap conductance larger than the normal conductance at any energy, similar to the re-entrance effect. This excess conductance is not quantitatively understood theoretically. A possible heating effect at finite bias, not present in the theories, can prevent the direct comparison.

\par We would like to acknowledge B. Pannetier, A.F. Volkov, A.D. Zaikin, F.W.J. Hekking, F. Pistolesi and J.C. Cuevas for valuable discussions and J.L. Thomassin for technical support.

\end{document}